\pdfoutput=1
\documentclass[floats,floatfix,showpacs,preprint,prd,superscriptaddress,onecolumn]{revtex4}
\begin{document}

\title{Direct Smooth Reconstruction of Inflationary Models in $f(R)$ Gravity }

\author{Gianluca Giacomozzi}
\email{gianluca.g.giacomozzi@gmail.com }
\author{Sergio Zerbini}
\email{sergio.zerbini@unitn.it}
\affiliation{Dipartimento di Fisica, Universit\`a di Trento,\\Via Sommarive 14,38123 Povo (TN), Italy}


\begin{abstract}
\noindent 
Starting from the parametrization of the spectrum of scalar perturbations generated during inflation in terms of the number of $e$--folds $N$ and using the approach recently developed by Starobinsky,  dubbed ``direct smooth reconstruction'', we show that, in the slow--roll approximation, it is possible to reconstruct the inflaton potential in the Einstein frame  and its corresponding $f(R)$ gravity Lagrangian model.
Viable inflationary models in agreement with the latest observational data released by the Planck and the BICEP2/Keck Array collaborations are recovered.
Furthermore, we show that, under some reasonable assumptions, the Starobinsky method is capable of describing consistently power law inflation, $R+R^2$ (Starobinsky) inflation, quasi--invariant inflation  and $\alpha$--attractors inflation.\end{abstract}



 \maketitle


\section{Introduction}

Modern cosmology is based on the idea that there was an epoch in the very early Universe, called inflation, characterized by a sufficiently long period of accelerated expansion. 
Inflation (see \cite{LIDDLE1,MUK,Odin23} for reviews) was introduced several years ago  \cite{Staro79,Staro,Guth,linde,Al} (see also  \cite{brout,Kaza,Sato}) in the attempt to solve some problems present in the Standard Big Bang cosmological model.
In particular, thanks to the inflationary paradigm, the spacetime, on large scales, naturally becomes flat, homogeneous and isotropic as we can observe today.
Furthermore, fluctuations of the related fields may be considered at quantum level. Thus, the inflation mechanism gives, in a quite natural way, an explanation of the tiny  anisotropies in the CMB spectrum and in the observed galactic structures in agreement with the results released by Planck and the \linebreak BICEP2/Keck–Array collaboration \cite{PLANCK,bicepkeck}.

Even if the current observational data set rigid constraints on the viability of possible models, in the literature one may find several inflationary models. Most of them are based on General Relativity (GR) with an additional minimally coupled scalar field, their properties being strictly dependent on the form of the potential.

However there exists another approach that can explain the acceleration in cosmology framework.
This corresponds to make use of the modified gravity scenario. Here GR  is modified and instead of Einstein--Hilbert action one considers as gravitational action a generic smooth function $f(R)$ of the Ricci scalar. This approach may be justified by invoking  quantum gravitational corrections. Within this approach, a very successful and  simple  model is the well known Starobinsky model \cite{Staro,Staro1} where  a quadratic term in the Ricci scalar is added to the usual Einstein--Hilbert action.

In this paper, we would like to make use of  a reconstruction method, due to Alexei starobinsky and dubbed ``direct smooth reconstruction'' \cite{Staro-Gui,Star-flag,Gian}. The main idea is to reconstruct $f(R)$ starting from the power spectrum parametrized by the e--fold number $N$. In the case of  standard inflation model, see for example  \cite{Hodges}.

Recently another method of reconstruction has been presented in \cite{VM1,VM2}, where the starting point is a parametrization of the barotropic index in terms of $N$. We will discuss how these two reconstruction methods are related, and furthermore their connection with the inflationary models  obtained in the $\alpha$--attractors formalism \cite{KL,KL2,KLR,GKLR} and within
the so called quasi--scale invariant approach \cite{RVZV,Ka,Ka1}.

Several other papers dealing with reconstruction procedures have been appeared. An incomplete list includes \cite{Cho,Cho1}, \cite{Odintsov1} and references therein.

The paper is organized as follows: in Section 2 is presented a short review of inflation in scalar theories and in $f(R)$ gravity.
In Section 3, we discuss the direct smooth  reconstruction for both the potential $V(\phi)$ and the corresponding $f(R)$ scenario starting for a suitable parametrization of the power spectrum. Furthermore, we show the equivalence between the direct smooth reconstruction method and the Mukhanov one.
Making use of direct smooth reconstruction, in Section 4 and 5 we investigate in details two special cases which generate different predictions for the spectral indexes and we show that, in these particular frameworks, the corresponding reconstructed models have, respectively, the same properties of several models of inflation and $\alpha$--attractors.
In Section 6, the conclusions are presented. 

Throughout the paper, we use units of $c=\hbar=1$ and we define $\kappa^2 = 8 \pi G = 8 \pi /m_P^2$ where $G$ is the gravitational constant and $m_P = 1.22\times 10^{19}$ GeV is the Planck mass.


\section{Inflation: general features}

In this section, we revisit some features of inflation and in particular the ones typical in the contexts of scalar theories and $f(R)$ theories of gravity. 


\subsection{Scalar inflation} 

To begin with, we recall the flat Friedman--Lema\^{i}tre--Robertson--Walker (FLRW) metric:
\begin{equation}
	\label{FLRW}
	ds^2 = - dt^2 + a^2(t) \left[dr^2 +r^2 \left( d \vartheta^2 + \sin^2 \vartheta d \varphi^2 \right) \right],
\end{equation}
where $a(t) \equiv a$ is the scale factor of the Universe at cosmological time $t$.
The minimally coupled scalar inflaton action reads:
\begin{equation}
\label{action scalar theory}
	S_E = \int d^4 x \, \sqrt{-g} \left[\frac{1}{2 \kappa^2} R - \frac{1}{2} (\partial \phi)^2 - V(\phi)  \right].
\end{equation}
The related energy density $\rho$ and pressure $ p$ are:
\begin{equation}
	\rho = \frac{\dot{\phi}^2}{2} + V(\phi), \qquad p = \frac{\dot{\phi}^2}{2} - V(\phi),
\end{equation}
and the Friedmann equations:
\begin{equation}
	\label{friedmann eqns scalar field}
	H^2 = \frac{\kappa^2}{3} \left(\frac{\dot{\phi}^2}{2} + V(\phi)\right), \qquad \dot{H} = - \frac{\kappa^2}{2} \dot{\phi}^2. 
\end{equation}
The barotropic index is:
\begin{equation}
  \label{omega}
  \omega= \frac{p}{\rho} =\frac{ \dot{\phi}^2-2V(\phi)}{ \dot{\phi}^2+2V(\phi)}. 
\end{equation}

The inflationary (quasi--De Sitter) expansion  is recovered  in the slow--roll approximation, namely  if $\dot{\phi} \ll V(\phi)$ and $|\ddot{\phi}| \ll 3 H \dot{\phi}$.
If we introduce the slow--roll parameters or  Hubble flow functions:
\begin{equation}
	\label{def slow--roll parameters}
	\epsilon_1 = -\frac{\dot{H}}{H^2}, \qquad \epsilon_2 = \frac{\dot{\epsilon_1}}{H \epsilon_1},
\end{equation}
then inflation takes place as soon as these quantities remain very small \emph{i.e.} $\{\epsilon_1, |\epsilon_2|\} \ll 1$ while it ends when they become of order of unity \emph{i.e.} $\{\epsilon_1, |\epsilon_2|\} \sim \mathcal O(1) $.
In particular, in the slow--roll approximation the first Friedmann equation  and the equation of motion for $\phi$ become:
\begin{equation}
	\label{slow--roll friedmann, continuity}
	H^2 \sim \frac{\kappa^2}{3} V(\phi), \qquad \dot{\phi} \sim - \frac{V'(\phi)}{3H} ,
\end{equation}
where the prime denotes the derivative with respect to the argument.

The number of $e$--folds left to the end of inflation is:
\begin{equation}
	\label{N}
	N = \log \left[ \frac{a(t_f)}{a(t)}  \right] = \int_t^{t_f} H \,dt \sim \kappa^2 \int_{\phi_f}^\phi \frac{V}{V'} \,d \phi, 
\end{equation}
where we have used the slow--roll result (\ref{slow--roll friedmann, continuity}) and we have defined $a(t_f)$ as the scale factor at the end of inflation with $t_f$ the related time.
In order to have the thermalization of the observable Universe is required that the total number of $e$--folds, which gives the total amount of inflation:
\begin{equation}
	\mathcal N = N_{t=t_i} = \log \left[ \frac{a(t_f)}{a(t_i)}  \right],
\end{equation}
must be $50 \lesssim \mathcal N \lesssim 60$ having defined $a(t_i)$ as the scale factor at the beginning of inflation with $t_i$ the related time.


\subsection{\boldmath Inflation in $f(R)$ gravity \unboldmath}

As well known, an alternative description of the early--time (or current) acceleration may be achieved  by the use of $f(R)$ modified theories of gravity \cite{So,TSU,No,Ca,Noji}, where the Einstein--Hilbert term in the (Jordan frame) action of GR gets replaced by an arbitrary smooth function $f(R)$ of the Ricci scalar R:
\begin{equation}
	\label{action jordan}
	S_J = \frac{1}{2 \kappa^2} \int d^4x \, \sqrt{-g} f(R).
\end{equation}

A crucial assumption in $f(R)$ gravity is that $f''(R)$ is not vanishing. As a result, there is an additional  scalar degree of freedom, dubbed scalaron, proportional to
$ f'(R)=\frac{d f}{d R}$. In fact, making use of  the following metric redefinition:
\begin{equation}
	\label{conf transf}
	\tilde{g}_{\mu \nu} = \Omega^2 g_{\mu \nu}, 	
\end{equation} 
where $\Omega^2$: 
\begin{equation}
	\Omega^2 = F(R) \equiv f'(R),
\end{equation}
the action (\ref{action jordan}) gets transformed into the Einstein frame action:
\begin{equation}
	\label{action Einstein frame}
	S_E = \int d^4x\, \sqrt{-\tilde{g}} \left[\frac{1}{2 \kappa^2} \tilde{R} - \frac{1}{2} (\tilde{\partial} \phi)^2 - V(\phi)  \right],
\end{equation}
 the scalar field $\phi$ is: 
\begin{equation}
	\label{phi einstein frame}
	 \kappa \phi = \sqrt{\frac{3}{2}} \log F\,,
\end{equation}
and  the related potential is:
\begin{equation}
	\label{potential Einstein frame in term of f}
	V(\phi) =  \frac{FR-f}{2 \kappa^2 F^2}\,.
\end{equation}

As already mentioned, one  successful $f(R)$ gravity inflation model is  the Starobinksy model defined by:
\begin{equation}
	\label{Star model}
	f(R) = R + \frac{R^2}{6M^2}\,.  	
\end{equation} 
The related   scalar potential in the Einstein frame is:
\begin{equation}
\label{star model potential}
	V(\phi) = \frac{3 M^2}{4 \kappa^2} \left( 1- e^{-\sqrt{2/3} \, \kappa \phi} \right)^2.	
\end{equation}


\section{Starobinsky Reconstruction  Method}

In this Section, we introduce the Starobinsky direct smooth reconstruction method. We recall that in the slow--roll approximation, the spectrum  $\mathcal P_{\mathcal R}$ of scalar perturbation, after inflation are simply related to the properties of the potential $V(\phi)$ at the  horizon exit \cite{Hawk82,Staro82,GuPi}.
As a matter of fact, at the lowest order, one has  \cite{scalar}:
\begin{equation}
	\label{scalar perturbation}
	\mathcal P_{\mathcal R}(k) = \frac{H^4}{4 \pi^2 \dot{\phi}^2} \sim \frac{\kappa^6}{12 \pi^2} \frac{V^3}{V'^2}\,.	
\end{equation} 
Here is left understood that all the quantities are evaluated at the horizon crossing, the associated comoving wave number being $k=Ha$, again at horizon crossing.

The key concept at the base of the direct smooth  reconstruction is the ``transformation of the power spectrum into a potential'' \cite{Hodges}.
 
We can re--parametrize the spectrum of scalar perturbations in terms of the number of $e$--folds $N$ in the following way:
\begin{equation}
	\label{v' su v}
	\frac{V'}{V^{3/2}} \, d \phi = \sqrt{\frac{\kappa^6}{12 \pi^2} } \frac{1}{\mathcal P_{\mathcal R}(N)^{1/2}} \frac{d \phi(N)}{d N} \, d N.
\end{equation}
From (\ref{N}) we have that:
\begin{equation}
\label{sigma = N}
	\frac{d \phi}{d N}  = \frac{1}{\kappa^2}  \frac{V'}{V} \sim \frac{1}{\kappa^2} \sqrt{\frac{\kappa^6}{12 \pi^2} } \frac{V^{1/2}}{\mathcal P_{\mathcal R}(N)^{1/2}}.
\end{equation}  
If we now substitute (\ref{sigma = N}) into (\ref{v' su v}) and multiply both sides of the equation by $V^{-1/2}$, then the right--hand side becomes independent on the potential and therefore can be integrated to:
\begin{equation}
\label{V definito}
	\frac{1}{V(N)}  = \frac{1}{V_0} - \frac{\kappa^4}{12 \pi^2} \int_{N_0}^N \frac{1}{\mathcal P_{\mathcal R}(N)} \, dN,   
\end{equation}
where $V_0^{-1}$ is an integration constant. 
Thus, the expression of $V(\phi)$ is  parametrized by $N$ while $\phi(N)$ is obtained from integrating (\ref{sigma = N}) using the above definition for the potential:
\begin{equation}
\label{kappa phi}
	\kappa \phi(N) = \kappa \phi_0 + \int_{N_0}^{N} \sqrt{\frac{d}{dN} \left(\log \frac{V(N)}{V_*}\right) } \, dN.
\end{equation}
having introduced the reference quantity $V_*$ in order to make the argument of the logarithm dimensionless.

For any given possible parametrization $\mathcal P_{\mathcal R}(N)$, the form of $V(\phi)$ can be reconstructed by solving (\ref{V definito}) and (\ref{kappa phi}), subsequently the expression for $\phi(N)$ must be inverted in $N$ and finally substituted back into (\ref{V definito}).

As a result, this reconstruction technique can be used for the determination of an $f(R)$ making use of
(\ref{potential Einstein frame in term of f}).

However, the presence of $F$ in (\ref{potential Einstein frame in term of f}) is problematic: we cannot isolate $f$ and write it as a function of the potential $V$ only.
In order to overcome this difficulty, we make the following assumption \cite{Staro-Gui, Star-flag}: let us factorize the modified Lagrangian as:
\begin{equation}
	\label{f R volte A}
	f(R) \equiv R^2 A(R)\,.
\end{equation}
and assume that  $A(R)$ is a slowly--varying function in $R$, namely:
\begin{equation}
	\label{A slowly v}
	\big| A'(R)\big| \ll \frac{A(R)}{R}, \qquad \big|A''(R)\big| \ll \frac{A(R)}{R^2}.
\end{equation}

These conditions are necessary to have {\em any} slow--roll inflation in $f(R)$ gravity, being the analogues of the two slow--roll conditions for the inflaton within GR. 
In fact, in reference \cite{Appleby} it has been explicitly shown that inflation occurs as soon as $f(R)$ is close to $R^2$ over some range of $R$. Furthermore, small--field (``new'') inflation in GR corresponds to the case when this range occurs only around one value $R=R_0$, while analogue of large--field (``chaotic'') inflation in GR takes place if the conditions (\ref{A slowly v}) are satisfied for some extended range of $R$, including
$R \to \infty$.

As a consequence, from (\ref{potential Einstein frame in term of f}) we get:
\begin{equation}
	\label{da V a A}
	V \sim \frac{1}{8 \kappa^2 A},
\end{equation}
and  (\ref{V definito}) translates into:
\begin{equation}
	\label{A}
 	A(N) = \frac{1}{8 \kappa^2 V_0}  - \frac{\kappa^2}{96 \pi^2} \int_{N_0}^N \frac{1}{\mathcal P_{\mathcal R}(N)} \, dN.
\end{equation} 
On the other hand, with regard to the second reconstruction formula (\ref{kappa phi}), we remind that, in the Einstein frame, the field $\phi$ is given by (\ref{phi einstein frame}). Thus, we have:
\begin{equation}
	\kappa \phi \sim \sqrt{\frac{3}{2}} \log \left( 2 R A \right),
\end{equation}
and:
\begin{equation}
\label{log2RA}
	\log \left( 2 R A \right) = C_0 + \int_{N_0}^N \sqrt{-\frac{2}{3} \frac{d}{dN} \left( \log \frac{A(N)}{A_*} \right)} \, dN ,
\end{equation}
where we have introduced a new dimensionless constant $C_0$ and a reference $A_*$ to make the argument of the logarithm dimensionless.

Since $A(R)$ is a slowly-varying function, this expression can be further simplified, and one may write $\log A\simeq \log \frac{1}{R_0}$,  $R_0$ constant. Thus, one has 
\begin{equation}
\label{logR}
	\log \left( \frac{R}{R_0} \right) =  \int_{N_0}^N \sqrt{-\frac{2}{3} \frac{d}{dN} \left( \log \frac{A(N)}{A_*} \right)} \, dN ,
\end{equation}

The $R$--dependence of the slowly--varying function $A$ then can be recovered following the same steps for the reconstruction of the potential $V(\phi)$ presented above.
Finally the $f(R)$ model is obtain just multiplying the function $A(R)$ by the term $R^2$. 

Since at the beginning of this approach we have used the slow--roll approximation in order to write $\mathcal P_{\mathcal R}$ as a function of $V$ and its derivative, the validity  of the direct smooth reconstruction, both in the potential reconstruction and in the $f(R)$ reconstruction, is guaranteed if the potential fulfills the slow--roll conditions:
\begin{equation}
	\frac{1}{2 \kappa^2} \left(\frac{V'}{V}\right) \ll 1, \qquad \bigg| \frac{V''}{\kappa^2 V}  \bigg| \ll 1.
\end{equation}
which, using (\ref{scalar perturbation}) and (\ref{sigma = N}), read:
\begin{equation}
	\label{consinstency}
	\frac{\kappa^4}{24 \pi^2} \frac{V}{\mathcal P_\mathcal{R}} \ll 1, \qquad \bigg| - \frac{1}{2} \frac{d}{dN} \log \mathcal P_\mathcal{R} + \frac{\kappa^4}{8 \pi^2} \frac{V}{\mathcal P_\mathcal{R}}  \bigg| \ll 1  
\end{equation}

Now, let us make a quite simple choice for the spectrum of  scalar perturbations, namely:
\begin{equation}
\label{ansatz P mono}
	 	\mathcal P_{\mathcal R} (N) = \mathcal P_0 \,(N+1)^{\delta+1},
\end{equation}
where $\delta$ is a free parameter.
With this ansatz, the slow--roll conditions (\ref{consinstency}) are both satisfied if:
\begin{equation}
	\label{delta slow-roll}
	|\delta| \ll 2N \sim 100,
\end{equation}
and therefore the parameter $\delta$ must be of order of unity.

With the above spectral ansatz, the potential reads
\begin{equation}
 \label{V N delta 0}
 V(N) = \left(\frac{1}{V_0} + \frac{\kappa^4}{12 \delta \pi^2 \mathcal P_0}
          \left((N+1)^{-\delta}-(N_0+1)^{-\delta}\right) \right)^{-1}\,.
\end{equation}
This leads to
\begin{equation}
  A(N)=\frac{1}{8 \kappa^2 V_0} + \frac{\kappa^2}{96 \delta \pi^2 \mathcal P_0}
          \left((N+1)^{-\delta}-(N_0+1)^{-\delta}\right)\,,
\label{A}
\end{equation}
and
\begin{eqnarray}
\label{log2RAA}
&&\log \left( \frac{R}{R_0} \right) =   \\ \nonumber
&+& \frac{\kappa}{12 \pi \sqrt{ \mathcal P_0}} \int_{N_0}^N
\frac{(N+1)^{-\frac{\delta+1}{2}}\, dN}{ \left(\frac{1}{8 \kappa^2 V_0} + \frac{\kappa^2}{96 \delta \pi^2 \mathcal P_0}
          \left((N+1)^{-\delta}-(N_0+1)^{-\delta}\right)\right)^{\frac{1}{2}}  }\,.
\end{eqnarray}

\subsection{Relation between the Starobinsky  and Mukhanov methods of reconstruction}

Before going into the effective calculation of the inflationary models that we can reconstruct, both in the scalar and $f(R)$ frameworks,  we will show that the direct smooth reconstruction may be related to the Mukhanov method of reconstruction developed by Mukhanov himself in \cite{VM1,VM2} and applied to $f(R)$ models in \cite{MSZ,SM}. However, the main difference between direct smooth  approach and Mukhanov one is that the first one does not consider different forms of $\mathcal P_{\mathcal R} (N) $ case by case and does not introduce additional quantities like EoS parameter $\omega$, but starts from the general expressions (\ref{A}) and (\ref{log2RA}) giving $f(R)$ for any measured $\mathcal P_\mathcal{R} $.
As a consequence,  ansatz like (\ref{ansatz P mono}) goes after that particular application to compensate the absence of data for  $\mathcal P_\mathcal{R}$ at all scales, namely $ 0 < N < 60$.

In order to solve the graceful exit problem from inflation, Mukhanov considered a sort of ``decaying cosmological constant'' whose (effective) EoS parameter must have, at the beginning of inflation, small, but non--vanishing, deviations with respect to the one of the standard  cosmological constant (\emph{i.e.} $1+\omega \sim 0 $) in order to provide a necessary duration of the accelerated expansion. 
The graceful exit from inflation is recovered, on the other hand, when this EoS becomes of order of unity (\emph{i.e.} $1+\omega \sim \mathcal O(1) $).
In particular, in order to describe the evolution of the barotropic index $\omega$, the number of $e$--folds $N$ left to the end of inflation is set as a time parameter.

Parametrizing $\omega$ in terms of $N$ however is totally equivalent to parametrize in terms of $N$ the spectrum of scalar perturbations $\mathcal P_{\mathcal R}$.
In order to see this, it is easy to show that the two Friedmann equations, when a scalar field is present, lead to:
\begin{equation}
	\epsilon_1 = \frac{3 (1+\omega)}{2}\,. 
\end{equation}
Furthermore,  making use of  (\ref{scalar perturbation}), with the aid of (\ref{friedmann eqns scalar field}), we find:
\begin{equation}
	\label{e1 star con V}
	\epsilon_1 = \frac{\kappa^4}{24 \pi^2} \, \frac{V}{ \mathcal P_{\mathcal R}}\,. 
\end{equation}
Thus, the above equation and (\ref{V definito}) give: 
\begin{equation}
	\label{equivalence star muk}
\frac{d}{d N}\left((1+\omega) \mathcal P_{\mathcal R}  \right)=3 (1+\omega)^2 \mathcal P_{\mathcal R}.
\end{equation}
The solution of above differential equation is:
\begin{equation}
	1 +\omega = -\frac{1}{3  \mathcal P_{\mathcal R} \int dN  (\mathcal P_{\mathcal R})^{-1}  }\,.  	
\end{equation} 
As a result, the power law parametrization (\ref{ansatz P mono}) corresponds to the following EoS parameter:
\begin{equation}
	1 +\omega = \frac{\delta}{3(N+1)},  	
\end{equation} 
which behaves as desired at the beginning of inflation \linebreak ($N \gg 1$) and also at the end of it ($N=0$) if $\delta$ is of order of unity, as we already know.

In particular, this parametrization of the barotropic index is exactly the one considered in \cite{MSZ,SM} which leads, for $\delta=1$, to the following inflationary model:
\begin{equation}
	f(R) = R + \frac{R^2}{6 \kappa^2 \rho_0} + \frac{\kappa^2 \rho_0}{6},  
\end{equation}
which corresponds to an extension of the Starobinsky model caused by the presence of an additive cosmological constant term where $\rho_0$ is an integration constant with dimensions \linebreak $[\rho_0]=M^4$.

\section{Spectral index and  tensor--to--scalar ratio }

Starting from the power law ansatz (\ref{ansatz P mono}), the viability of any given reconstructed model must be checked by comparing its predictions for the spectral index $n_s$ and the tensor--to--scalar ratio $r$ with the latest observational data coming from the Planck $2015$ satellite experiment \cite{PLANCK} combined with the ones of the BICEP2/Keck-Array collaboration \cite{bicepkeck}:
\begin{equation}
	\label{data}
	n_s = 0.968 \pm 0.006 \text{  (68 \% CL)}, \quad r < 0.07 \text{  (95 \% CL)}. 	
\end{equation} 
In the Einstein frame, at leading order, the spectral index and the tensor--to--scalar ratio are given by:
\begin{equation}
	n_s = 1 - 2 \epsilon_1|_{N} - \epsilon_2|_{N}, \quad r = 16 \epsilon_1|_{N}.
\end{equation}
We recall  the expression for the first  slow--roll parameter
\begin{equation}
	\label{e1 star con V}
	\epsilon_1 = \frac{\kappa^4}{24 \pi^2} \, \frac{V}{ \mathcal P_{\mathcal R}}, 
\end{equation}
while the second slow--roll parameters $\epsilon_2$ in (\ref{def slow--roll parameters}) can be written as only a function of $\epsilon_1$ if we transform the time derivative into a derivative with respect to the number of $e$--folds \emph{i.e.} $d/dt = - H d/dN$, obtaining finally:
\begin{equation}
	\label{e2}
	\epsilon_2 = - \frac{1}{\epsilon_1} \frac{d \epsilon_1}{d N} = - \frac{d}{d N} \log \epsilon_1.
\end{equation}
Therefore, using the above formulas with the expression of the potential given by
(\ref{V definito}) and the power law ansatz (\ref{ansatz P mono}), we get:
\begin{equation}
	\label{ns star}
	n_s = 1 - \frac{\delta+1}{ N +1},\\[.5em]
\end{equation}
\begin{equation}
	\label{r star}
	r = \frac{8 \delta}{( N +1) \left[1 + \frac{12 \pi^2 \delta \mathcal P_0 ( N +1)^{\delta}}{V_0 \kappa^4} -\frac{( N +1)^{\delta}}{(N_0 +1)^{\delta}}\right]}.\\[.5em]
\end{equation}
As we can see from the above expressions, the spectral index $n_s$ depends only on the free parameter $\delta$ that enters in the parametrization of the scalar perturbations as the exponent of the number of $e$--folds $N$.

As a simple check, the particular choice $\delta = -1$, which corresponds to a constant scalar spectrum, reduces to $n_s = 1$ namely to a scale invariant or Harrison--Zel'dovich spectrum \cite{HZS}.

On the other hand,  the tensor--to--scalar ratio depends also on the arbitrary integration constant $V_0^{-1}$ presents in the reconstruction formulas; therefore by setting in a suitable way $V_0^{-1}$ we can tune the $ N$--dependence of $r$.

We consider two cases:
\begin{eqnarray}
	\label{Vo caso 1}
		\text{case A:}& \quad  V_0^{-1} = \frac{\kappa^4}{12 \pi^2 \delta \, \mathcal P_0 (N_0 +1)^\delta}\,,\\[0.5em]
	\label{Vo caso 2}
		\text{case B:}& \quad V_0^{-1} = \frac{\kappa^4}{6 \pi^2 \delta \, \mathcal P_0 (N_0 +1)^\delta}\,.
\end{eqnarray}
which lead to the following expressions:
\begin{equation}
	\label{r caso 1}
	\mbox{case A:}\quad   r =  \frac{8 \delta}{ N +1} \,.
\end{equation}
\begin{equation}
\label{r caso 2}
\mbox{case B:}\quad   r = \frac{8 \delta}{( N +1) \left[ 1 + \frac{( N +1)^\delta}{(N_0 +1)^\delta} \right]} \sim    
       \frac{8 \delta (N_0 +1)^\delta}{( N +1)^{\delta+1}} \,,\quad  \mbox{$ N_0 \ll 1 $} \,,
\end{equation}       
\begin{equation}
\mbox{case B:} \quad  r  \sim \frac{8 \delta}{ N +1}\,, \quad  \mbox{$ N_0 \gg 1$} 
        \, .
\end{equation}
Thus, in the case B, depending on the value of the parameter $N_0$, we get two possible behaviours for the tensor--to--scalar ratio: linear in $1/ N$ in the limiting case $N_0 \gg 1$ while going like a generic power $1/ N^{\delta+1}$ in the other case when $N_0 \ll 1$.

%

\subsection{Inflationary attractors: case $A$  }

First of all, let us consider the case A equation (\ref{Vo caso 1}) for $V_0^{-1}$ for which we have :
\begin{equation}
	n_s = 1 - \frac{\delta +1}{\mathcal N +1}, \quad r = \frac{8 \delta}{\mathcal N +1}. 
\end{equation}
First, we note that for the usual range $ 50 < N< 60$, there is no a good agreement with Planck data. Only setting $ N = 45$, these two quantities are compatible with the observational data (\ref{data}) if:
\begin{equation}
\label{intervalli delta P mono}
	\text{for $n_s$:} \;\;\; 0.20 < \delta < 0.75, \qquad \text{for $r$:} \;\;\;  \delta < 0.40,
\end{equation}
confirming the relation (\ref{delta slow-roll}).
Moreover, since these two intervals are not disjoint, starting from the monomial ansatz, it is possible to reconstruct viable models for inflation.

With regard to the potential reconstruction, in this particular case, the term coming from the integration of the scalar perturbations evaluated at $N_0$ in (\ref{V definito}) for our ansatz (\ref{ansatz P mono}) exactly cancel the \emph{ad hoc} choice of the integration constant $V_0^{-1}$ allowing us to write the potential $V(N)$ as an indefinite integral of $\mathcal P_{\mathcal R}$ in the number of $e$--folds:
\begin{eqnarray}
	\label{v indefinito}
	V(N) 	&= \left( - \frac{\kappa^4}{12 \pi^2}  \int \frac{1}{\mathcal P_0 (N+1)^{\delta+1}} \, dN \right)^{-1} \\[.5em]
			&= \frac{12 \pi ^2 \delta \,\mathcal P_0 (N+1)^{\delta}}{\kappa^4}.
\end{eqnarray}
Plugging this expression into (\ref{kappa phi}) and then following the reconstruction procedure, we obtain the inflationary potential in the Einstein frame:
\begin{equation}
  V(\phi)=\frac{12 \pi^2 \delta \mathcal P_0 }{\kappa^4}
  \left(\frac{\kappa^2}{4 \delta} (\phi-\phi_0) \right)^{2 \delta},
\end{equation}
which behaves as a generic  power--law potential in the scalar field $\phi$. 

\indent In the $f(R)$ framework, on the other hand, we have from (\ref{A}):
\begin{equation}
	\label{A P mono}
	A(N) = \frac{\kappa^2}{96 \pi^2 \mathcal \delta \, P_0 (N+1)^\delta}, 
\end{equation}
and from (\ref{logR}):
\begin{equation}
	\log  \frac{R}{ R_0 } = \sqrt{\frac{2 \delta (N+1)}{3}},
\label{nzero}
\end{equation}
where we have absorbed the constants of integration. Combining the above results, one has
\begin{equation}
	A(R) = (\log  \frac{R}{ R_0 })^{-2\delta}\,.
\label{nzeroa}
\end{equation}
Here one should note that the related $A(R)$ is a decreasing function, namely with negative first derivative. Absorbing all
the constant of integration leads to
\begin{equation}
\label{f log in star}
	f(R) \sim \frac{R^2}{\left[ \log \left( \frac{R}{\mu^2}  \right) \right]^{2 \delta}}.
\end{equation}

As a result,  we have obtained a logarithmic correction to the pure scale invariant $R^2$ scenario.


We conclude this section with some remarks. Within the case A,  we have been  then able to generate in the Einstein frame, the power--law potential typical of large--field inflation \cite{LINCHAOS} with the same predictions of the spectral indexes $n_s$ and the tensor--to--scalar ratio $r$.
In particular, for the allowed values of $\delta$ (\ref{intervalli delta P mono}), the reconstructed model presents a modified convex--concave divide in the $(n_s,r)$--plane:
\begin{equation}
	\label{modified divide}
	r = \frac{8 \delta}{\delta +1}(1-n_s), 
\end{equation}
which is lower than the standard one, which is defined for $\delta = 1/2$ and corresponds to a linear potential in the scalar field $\phi$, matching therefore in a better way the observational data.

The same behaviour can be found also for the reconstructed $f(R)$ model.
Indeed, following the procedure developed in \cite{RVZ23}, we find that at leading order for small $r$ the model (\ref{f log in star}) predicts:
\begin{equation}
	1 - n_s = \frac{(\delta + 1)r}{8 \delta} + \mathcal O(r^{3/2}),
\end{equation}
independently on the values of the constant $C_1$ and on the mass scale $\mu$.

Under the assumption that $C_1 < 0$ and $\delta=1/2$, our $f(R)$ inflationary model is equivalent to the deformed quadratic model:
\begin{equation}
	f(R) \sim \frac{R^2}{ 1 + \gamma \log \left( \frac{R}{\mu^2}  \right) } 
\end{equation}
proposed in \cite{RVZ23,RVZV} which generates the standard convex--concave divide.
In analogy with QCD, the above model can be seen as a resummation of loop--corrections
 as already pointed out in \cite{RIN} and corresponds, on--shell in the slow--roll approximation, to the induced gravity model with the Coleman--Weinberg potential of the form:
\begin{equation}
\label{L p2 canonica}
		\mathcal{L}_J = \sqrt{-g} \left[ \xi \phi^2 R - \frac{1}{2} (\partial \phi)^2 - \lambda \phi^4 \left( 1+ \zeta_\gamma \log \left(\frac{\phi^2 }{\mu^2 }\right) \right) \right].
\end{equation}   
In this scenario, the reconstructed model (\ref{f log in star}) may be possibly originated by quantum loop--corrections whose resummation is controlled by the parameter $\delta$ which softly break the scale--invariance of the model introduced and investigated in \cite{Venturi,Turchetti}.


\section{Alpha--attractors: case $B$}

Let us now examine the case B, equation (\ref{Vo caso 2}).
Here, we have to consider separately the two regimes for the constant $N_0$ since now the tensor--to--scalar ratio depends on it. The $n_s$  index is the same
\begin{equation}
	n_s = 1 - \frac{\delta+1}{ N +1}\,.
\end{equation}
Instead for the other, one has 
\begin{equation}
r \sim  \frac{8 \delta (N_0 +1)^\delta}{( N +1)^{\delta+1}}  \quad  \mbox{$N_0 \ll 1$}\,,
\end{equation}
\begin{equation}
r \sim  \frac{8 \delta}{ N +1}  \quad  \mbox{$N_0 \gg 1$} \,.
\end{equation}
In the limit case $N_0 \gg 1$, the spectral indexes reduce to the ones found previously in the case A for $V_0^{-1}$.
This is due to the fact that the difference between the integration constant in the case A and B become smaller and smaller as soon as the value of $N_0$ increases.
Therefore, for large values of $N_0$, we obtain another time the predictions for the spectral indexes that are typical of
inflationary models within the case $A$.  

On the other hand, when $N_0 \ll 1$, the tensor--to--scalar ratio goes like $1/ N^{\delta+1}$ and therefore it is highly suppressed for the typical values of the number of $e$--folds.
Setting $\mathcal N = 60$, the spectral indexes are in agreement with the observational data  (\ref{data}) if:
\begin{equation}
	\text{for $n_s$:} \;\;\; 0.59 < \delta < 1.32, \qquad \text{for $r$:} \;\;\;  \delta > 0.
\end{equation}

\subsection{ Case $B$:  $\delta=1$}

Within the case $B$ and generic $\delta$, the direct smooth reconstruction has to deal with technical difficulties, and only $\delta=1$ and $\delta=1/2$ can be analytically treated.

Thus, first let us focus our attention  on the case $\delta=1$ since the last cosmological observations constrain the tensor--to--scalar ratio as $r \sim 1/ N^2$.
The potential  can be written as:
\begin{eqnarray}
	\label{V N delta 1}
	V(N) 	&= \left(\frac{\kappa^4}{6 \pi^2 \mathcal P_0 \, (N_0 +1)} - \frac{\kappa^4}{12 \pi^2}  \int_{N_0}^{N} \frac{1}{\mathcal P_0 \,(N+1)^2} \, dN \right)^{-1} \\[.5em]
			&= \frac{3 M^2}{4 \kappa^2} \left( 1+ \frac{N_0+1}{N+1}  \right)^{-1}\,,
\end{eqnarray}
where we have introduced a new mass:
\begin{equation}
	M^2 = \frac{16 \pi^2 \mathcal P_0 (N_0 +1)}{\kappa^2}. 
\end{equation}
With this form of the potential, once the equation (\ref{kappa phi}) is solved and then inverted in $N$, we are able to write:
\begin{equation}
	V(\phi) = \frac{3 M^2}{4 \kappa^2} \left[ \frac{1-(N_0+1) e^\frac{\kappa(\phi_0 - \phi)}{\sqrt{(N_0+1)}}}{1+(N_0+1) e^\frac{\kappa(\phi_0 - \phi)}{\sqrt{(N_0+1)}}}  \right]^2.
\end{equation}
Now, if we want that the above potential must be zero at $\phi=0$, we have to set:
\begin{equation}
	(N_0+1) e^\frac{\kappa \phi_0}{\sqrt{(N_0+1)}} = 1,
\end{equation}
from which we obtain finally:
\begin{equation}
\label{T model}
	V(\phi) = \frac{3 M^2}{4 \kappa^2} \tanh^2\left( \frac{\kappa \phi}{2 \sqrt{N_0 +1}}  \right).
\end{equation}

The corresponding $f(R)$ inflationary model can be found starting from the expression of the slowly--varying function:
\begin{equation}
	A(N) = \frac{1}{6M^2} \left( 1 + \frac{N_0+1}{N+1}  \right),
\end{equation}
Re--doing then all the steps in the reconstruction technique, in the limits $N \gg N_0$ and $R \gg R_0$, one gets (see \cite{Star-flag}):
\begin{equation}
	f(R) \sim \frac{R^2}{6 M^2} \left[ 1 + \left( \frac{R_0}{R}  \right)^{\sqrt{\frac{3}{2(N_0+1)} }} \right],
\end{equation}
which again behaves like a modification of the pure $R^2$ scenario but this time by means of a slowly--varying function that goes like $(1/R)^{\gamma}$.
In particular, the above model reduces exactly to the Starobinsky model (\ref{Star model}) for $R_0=6M^2$ and $N_0 = 1/2$.



We conclude this subsection observing that, in the specific case $\delta = 1$ and in the limit of small $N_0$, we are able to generate the predictions typical of $\alpha$--attractors formalism:
\begin{equation}
	n_s = 1 - \frac{2}{ N+1}, \quad r = \frac{12 \alpha}{( N+1)^2}, 
\end{equation}
if we make the following identification:
\begin{equation}
	\label{relation alpha No}
	8(N_0+1) = 12 \alpha.
\end{equation}
This fact is confirmed if we looked at the form of the reconstructed potential $V(\phi)$.
Using again the relation (\ref{relation alpha No}), indeed the potential (\ref{T model}) reduces to \cite{Star-flag}:
\begin{equation}
	V(\phi) = \frac{3 M^2}{4 \kappa^2} \tanh^2\left( \frac{\kappa \phi}{\sqrt{6 \alpha}}  \right),	
\end{equation} 
and this is related to  the so--called T--model \cite{KL,GKLR} inflation, namely the above identification chooses
one specific case among the all  $\alpha$--attractors. Here it is a simple consequence of the choice $\delta=1$.
This particular expression for the potential was originally introduced by Kallosh \emph{et al} since it is the simplest superconformally invariant theory with spontaneous symmetry breaking that leads to an inflationary expansion of the Universe.
The predictions of this model interpolate those of a large--field model with a quadratic potential (\emph{i.e.} $V(\phi) \sim \phi^2$) for $\alpha \gg 1$ and the $R+R^2$ model for $\alpha = 1$.

This property is perfectly consistent with our results, taking care of (\ref{relation alpha No}), as can we seen from (\ref{r caso 2}) in the $\delta=1$ case.

We stress that the power of the direct smooth  method of reconstruction lies in the fact that it shows manifestly the correspondence between the T--model potential and the associated $f(R)$ model of gravity for different values of $N_0$.
Indeed, as we have seen, the $N_0 = 1/2$ (\emph{i.e.} $\alpha=1$) case generates naturally the Starobinsky model.
On the other hand, in the $N_0 \gg 1$ case, the $f(R)$ model which corresponds to the quadratic potential in $\phi$ is the following one valid in the limit $R \gg R_0$ (see \cite{Star-flag}):
\begin{equation}
	f(R) = \frac{R^2}{6M^2} \left( \frac{1+ \left( \frac{R_0}{R}\right)^{\sqrt{\frac{3}{2(N_0+1)} }} }{1- \left( \frac{R_0}{R}\right)^{\sqrt{\frac{3}{2(N_0+1)} }}}  \right)^2,    	
\end{equation}   
that in the $N_0 \gg 1$  gives:
\begin{equation}
	f(R) \sim \frac{R^2}{\log^2\left( \frac{R}{R_0}  \right)^\gamma}\,,
\end{equation}
where $\gamma=\sqrt{\frac{3}{2(N_0+1)} }  $. Again we have obtained logarithmic correction to $R^2$ as in the case A for $V_0^{-1}$.
Furthermore, we note that the denominator of the above model does  resemble the one previously found in (\ref{f log in star}).

\subsection{\boldmath Case $B$ $\,,$   $\delta=\frac{1}{2}$ \unboldmath}
In this second subsection, we consider the case $B$, but with the choice $\delta=\frac{1}{2}$, since this case is also marginally admitted by data. One gets
\begin{equation}
\label{mezzo0}
  n_s = 1 - \frac{3}{2( N +1)}\,,
\end{equation}
\begin{equation}
r  \sim  \frac{4 (N_0 +1)^{1/2}}{( N +1)^{3/2}}  \quad  \mbox{$N_0 \ll 1$}\,, 
\end{equation}
\begin{equation}
r  \sim  \frac{ 4}{N+1}  \quad  \mbox{$N_0 \gg 1$}\,. 
       \,.
\end{equation}
 In the $(n_s,r)$--plane, one has
\begin{equation}
	\label{mezzo}
	r = \frac{8 (2N_0 +2)^{1/2} }{3\sqrt{3}}(1-n_s)^{3/2}\,. 
\end{equation}

The potential as a function of $N$ reads
\begin{equation}
	\label{V N delta 1/2}
V(N)= \left(\frac{\kappa^4}{3 \pi^2 \mathcal P_0 \,\sqrt{ (N_0 +1)}} - \frac{\kappa^4}{12 \pi^2}  \int_{N_0}^{N} \frac{dN}{\mathcal P_0 \,(N+1)^{3/2}} \right)^{-1}\,,
\end{equation}
namely
\begin{equation}
V(N)= \frac{3 M_{1/2}^2}{4 \kappa^2} \left( 1+ \sqrt{\frac{N_0+1}{N+1}}  \right)^{-1}\,.
\end{equation}
Here we have introduced the new mass:
\begin{equation}
	M_{1/2}^2 = \frac{8 \pi^2 \mathcal P_0 \sqrt{(N_0 +1)}}{\kappa^2}. 
\end{equation}
The other useful result is the expression for $\phi$, which can be exactly evaluated
\begin{equation}
\kappa(\phi-\phi_0)=\sqrt{8}(N_0+1)^{1/4}\left( \left[(N+1)^{1/2}+(N_0+1)^{1/2}\right]^{1/2} -\sqrt{2}(N_0+1)^{1/4}    \right)\,,  
\label{phi}
\end{equation}

The reconstruction of the potential leads to
\begin{equation}
  V(\phi)= \frac{3 M_{1/2}^2}{4 \kappa^2}
   \frac{\left(\kappa(\phi-\phi_0)+4\sqrt{N_0+1}  \right)^2 -8(N_0+1)}{\left(\kappa(\phi-\phi_0)+4\sqrt{N_0+1}  \right)^2 }\,. 
\end{equation}
We may fix the arbitrary constant $\phi_0$ requiring $V(0)=0$. As a result, in the above expression, one gets
 \begin{equation}
  \kappa \phi_0 =(4-2\sqrt{2})\sqrt{(N_0+1})\,.
 \end{equation}
 Thus,  taking into accout the expression for $M^2_{1/2} $, we have

\begin{equation}
\label{pot}
  V(\phi)= \frac{6 \pi^2  \mathcal P_0 }{\kappa^4}
  \left(\frac{2 \sqrt{8}(N_0+1)\kappa \phi+ \sqrt{(N_0+1)}\kappa^2 \phi^2}{\left(\kappa(\phi+\sqrt{8(N_0+1})  \right)^2  } \right)\, . 
\end{equation}
Let us discuss the  asymptotic limit $N_0 >>1$, namely
\begin{equation}
V(\phi) \simeq \frac{6  \pi^2  \mathcal P_0}{\sqrt{2} \kappa^3}\phi \,,
\end{equation}
namely a linear potential. This is in agreement with the related spectral asymptotics of the index $r$, and the reconstruction procedure is similar to the one discussed in the previous Section.

In the other limit  $N_0<< 1$, in the $(n_s,r)$--plane, one has
\begin{equation}
	\label{mezzo}
	r = \frac{8 (2N_0 +2)^{1/2} }{3\sqrt{3}}(1-n_s)^{3/2}\,, 
\end{equation}
and the potential may taken in the form (\ref{pot}). In this case, the potential in the Einstein frame is an increasing function of $\phi$, unbounded from below. As a check, we may discuss the slow-roll approximation,  considering the above potential  in the large $\phi$ approximation. Then, it is easy to check that one gets the relation (\ref{mezzo}).

The model is  compatible with Planck data: for $n_s=0.968$, one gets $ r=0.0124$ within a number of e-fold around $N=60$.

Finally, in order to reconstruct the related $f(R)=R^2A(R)$, one has to make use of the relation $\kappa(\phi-\phi_0)= \sqrt{\frac{3}{2}}\log\frac{R}{R_0} $ and relation (\ref{phi}),
obtaining 
\begin{equation}
\log(\frac{R}{R_0})=   \sqrt{\frac{16}{3}}(N_0+1)^{1/4}\left( \left[(N+1)^{1/2}+(N_0+1)^{1/2}\right]^{1/2} -\sqrt{2}(N_0+1)^{1/4}\right)    \,,
\end{equation}
 Then, in the approximation we are dealing with, we may write
\begin{equation}
\frac{(N_0+1)^{1/2}}{ (N+1)^{1/2}}=\frac{8 (N_0+1)}{16 (N_0+1)+ 16 \sqrt{\frac{3}{2}(N_0+1)}\log \frac{R}{R_0}+  3\log^2\frac{R}{R_0} } \,.
\end{equation}
As a result,  one gets

\begin{equation}
f(R)=\frac{R^2}{6 M_{1/2}^2} \left(1+\frac{16 (N_0+1)}{16 (N_0+1)+ 16 \sqrt{\frac{3}{2}(N_0+1) }\log \frac{R}{R_0}+  3\log^2\frac{R}{R_0} } \right)  \,.
\end{equation}
We conclude this Section observing that in the recent papers \cite{Odintsov20}, alpha-attractors within $f(R)$ modified gravitational models have been investigated with a different approach. In these paper, the authors have been implemented the reconstruction of inflationary $f(R)$ models working directly in the Jordan frame, namely, making use of the values of slow-roll parameters in this frame.

The novelty in the Starobinsky approach consists first in  dealing with the reconstruction of the potential in the Einstein frame, and then the reconstruction of the modified $f(R)$ models is achieved making use of  the well known standard procedure.

 \section{Conclusions}

In this paper, we have shown that the parametrization of the spectrum of scalar perturbations in terms of $N$, the number of $e$--folds left to the end of inflation, permits the reconstruction of the potential, present in the Einstein frame action, in function of the scalar field $\phi$.
Within the same framework, we have also shown that the direct reconstruction of $f(R)$ inflationary models is also possible under the assumption, which was proven in the aftermath, that these ones can be expressed as a modification of a pure $R^2$ model deformed by the presence of a slowly--varying function of $R$.

We have also shown that this reconstruction technique is connected to the Mukhanov one, since there is a one--to--one correspondence between the parametrization of $\mathcal P_{\mathcal R}$ and the one of the barotropic index $\omega$.
Taking advantage of this, we have shown that a power law form for $\mathcal P_{\mathcal R}$ corresponds to a ``decaying cosmological constant'' that solves the graceful exit problem.

Furthermore, we have studied the impact of the integration constant $V_0^{-1}$ on the reconstruction technique.
We have found that it controls the $\mathcal N$--dependence of the tensor--to--scalar ratio $r$ while it leaves unchanged the spectral index $n_s$.
In particular, it is possible to fix $V_0^{-1}$ in such a way that $r$ becomes linear in $1/\mathcal N$ (case A) or more generically behaves like $1/\mathcal N^{\delta+1}$ (case B).

In the first case, the predictions of the spectral indexes reproduce the ones typical of large--field inflation as revealed by the resolution of the reconstruction formulas for the potential $V(\phi)$ that can be written as a power in the scalar field.
The corresponding $f(R)$ model, which possesses logarithmic corrections, results to be a generalization of the quasi--scale invariant model presented in \cite{RVZ23,RVZV} with a  modified convex--concave divide in the $(n_s,r)$--plane that crosses the most interesting part of the observational data for $n_s$ and $r$. 
Such a model can possibly be originated by quantum loop--corrections whose resummation is controlled by the parameter $\delta$.
The feasibility of this model could be tested if future observations would confirm the existence of a ``lower'' concave--convex divide.
 
In the case B, on the other hand, we have focused our attention to the situation when, at $\delta=1$, the tensor--to--scalar ratio becomes quadratic in $1/\mathcal N$ as favored by the cosmological data.
Here we have found that the predictions of the spectral indexes resemble the ones present in the $\alpha$--attractors formalism if now the role of the free parameter $\alpha$ that labels the members of the class is played by $N_0$.
Furthermore, this result is confirmed by the reconstructed potential which takes the form of the one present in the T--model inflation which tends to a quadratic potential when $N_0 \gg 1$.
Just like $V(\phi)$, also the corresponding $f(R)$ model depends on the value of $N_0$.
In the large $N_0$ limit we have obtained again a logarithmic correction to the pure $R^2$ gravity in agreement with the results of the case A while, in the small $N_0$ limit, the $f(R)$ model gets corrected by a slowly--varying function in $R$ and reproduces exactly the Starobinsky model and its predictions for the choice $N_0=1/2$. Furthermore, we may add that for $\delta >1$, analogues of small-field inflation in $f(R)$ gravity are also possible.

Taking into account all of these results, we believe that the Starobinksy method of reconstruction can be considered one of the most powerful approaches for generating viable inflationary models thanks to its capability of ranging over different models.

\section{Acknowledgements}
This paper has been written in memory of Alexey Starobinsky, who prematurely  passed way in December 2023. We are indebted with him for usefull suggestions and kind support.
S.Z. also thanks Massimilano Rinaldi, Luciano Vanzo, Giovanni Venturi and Alexander Kamemshchik for discussions and suggestions.




\begin{thebibliography}{99} 

\bibitem{LIDDLE1} 
	A. R. Liddle, D. H. Lyth, 
	\emph{Cosmological Inflation and Large--Scale Structure}, Cambridge University Press (2000).

\bibitem{MUK} 
	V. Mukhanov, 
	\emph{Physical Foundations of Cosmology}, Cambridge University Press (2005).

\bibitem{Odin23}
S.~D.~Odintsov, V.~K.~Oikonomou, I.~Giannakoudi, F.~P.~Fronimos and E.~C.~Lymperiadou,
Symmetry \textbf{15} (2023) no.9, 1701.

\bibitem{Staro79}  
	A. A. Starobinsky,
  	\emph{JETP Lett.} \textbf{30}, 682 (1979).
  	[\emph{Pisma Zh.\ Eksp.\ Teor.\ Fiz.}  {\bf 30}, 719 (1979)].
  
\bibitem{Staro} 
	A. A. Starobinsky, 
	\emph{Phys. Lett. B} \textbf{91}, 99 (1980). 

\bibitem{Guth} 
	A. H. Guth, 
	\emph{Phys. Rev.} \textbf{23}, 347 (1981). 

\bibitem{linde} 
  	A. D. Linde,
  	\emph{Phys. Lett. B}  {\bf 108}, 389 (1982).

\bibitem{Al} 
  	A. Albrecht, P. J. Steinhardt,
  	\emph{Phys.\ Rev.\ Lett.}  {\bf 48}, 1220 (1982).

\bibitem{brout} 
  	R. Brout, F. Englert, E. Gunzig,
  	\emph{Annals Phys.}  {\bf 115}, 78 (1978).

\bibitem{Kaza} 
  	D. Kazanas,
  	\emph{Astrophys.\ J.}  {\bf 241}, L59 (1980).

        
\bibitem{Sato} 
	K. Sato, 
	\emph{Phys. Lett. B} \textbf{99}, 66 (1981). 


\bibitem{PLANCK} 
	P. A. R. Ade \emph{et al.} [Planck Collaboration]. 

\bibitem{bicepkeck} 
	P. A. R. Ade \emph{et al.} [BICEP2/Keck Array Collaboration], 
	\emph{Phys. Rev. Lett.} \textbf{116}, 031302 (2015). 

\bibitem{Staro1} 
	A. A. Starobinsky, 
	\emph{Sov. Astron. Lett. B} \textbf{9}, 302 (1983).

\bibitem{Staro-Gui}
	A. A. Starobinsky, invited talk at \emph{The  International Conference on Gravitation and Cosmology/The Fourth Galileo--Hu Guangqi Meeting}, 4--8 June, Beijing, China (2015),

\bibitem{Star-flag} 
	A. A. Starobinsky, Invited talk at \emph{2nd FLAG Meeting ``The Quantum and Gravity''}, 6--8 Jun, Trento, Italy (2016), 
\bibitem{Gian}
Gianluca Giacomozzi, ``Reconstrucion of Inflation Models in Modified Gravity Theories'', Master Thesis, Department of Physics,
University of Trento, Italy (2016).


\bibitem{Hodges} 
 	H. M. Hodges and G. R. Blumenthal,
 	\emph{Phys.\ Rev.\ D} {\bf 42}, 3329 (1990).

\bibitem{VM1} 
	V. Mukhanov, 
	\emph{Eur. Phys. J. C} \textbf{73}, 2486 (2013). 

\bibitem{VM2} 
	V. Mukhanov, 
	\emph{Fortsch. Phys.} \textbf{63}, 36 (2015). 

\bibitem{KL} 
	R. Kallosh, A. D. Linde, 
	\emph{Phys. Rev. D} \textbf{91}, 083528 (2015). 

\bibitem{KL2}  
	R. Kallosh, A. D. Linde, 
	\emph{JCAP} \textbf{1307}, 002 (2013). 

\bibitem{KLR} 
	R. Kallosh, A. D. Linde, D. Roest, 
	\emph{JHEP} \textbf{1311}, 198 (2013). 

\bibitem{GKLR} 
	M. Galante, R. Kallosh, A. D. Linde, D. Roest, 
	\emph{Phys. Rev. Lett.} \textbf{114}, 141302 (2015). 

\bibitem{RVZV} 
	M. Rinaldi, L. Vanzo, S. Zerbini, G. Venturi, 
	\emph{Phys. Rev. D} \textbf{93}, 024040 (2016). 

\bibitem{Ka} 
  	K. Kannike, G. Hütsi, L. Pizza, A. Racioppi, M. Raidal, A. Salvio, A. Strumia,
  	\emph{JHEP} {\bf 1505}, 065 (2015).

\bibitem{Ka1} 
 	K. Kannike, A. Racioppi, M. Raidal,
  	\emph{JHEP} {\bf 1406}, 154 (2014).


\bibitem{Cho}
S.~Choudhury and A.~Mazumdar,
``Reconstructing inflationary potential from BICEP2 and running of tensor modes,''
[arXiv:1403.5549 [hep-th]]:
S.~Choudhury,
Nucl. Phys. B \textbf{894} (2015), 29-55.
\bibitem{Cho1}
S.~Choudhury,
Phys. Dark Univ. \textbf{11} (2016), 16-48.

\bibitem{Odintsov1}
S.~D.~Odintsov and V.~K.~Oikonomou,
Nucl. Phys. B \textbf{929} (2018), 79-112;

S.~D.~Odintsov and V.~K.~Oikonomou,
Annals Phys. \textbf{388} (2018), 267-275.

S.~D.~Odintsov, V.~K.~Oikonomou and T.~Paul,
Nucl. Phys. B \textbf{959} (2020), 115159.


\bibitem{So} 
  	T. P. Sotiriou, V. Faraoni,
  	\emph{Rev.\ Mod.\ Phys.}\  {\bf 82}, 451 (2010).

\bibitem{TSU} 
	A. De Felice, S. Tsujikawa, 
	\emph{Living Rev. Rel.} \textbf{13}, 3 (2010). 

\bibitem{No} 
  	S. Nojiri, S. D. Odintsov,
  	\emph{Phys.\ Rept.}\  {\bf 505}, 59 (2011).
  
\bibitem{Ca} 
  	S. Capozziello, M. De Laurentis,
  	Phys.\ Rept.\  {\bf 509}, 167 (2011).


\bibitem{Noji}
S.~Nojiri, S.~D.~Odintsov and V.~K.~Oikonomou,
Phys. Rept. \textbf{692} (2017), 1-104.



\bibitem{Hawk82} 
  	S. W. Hawking,
  	\emph{Phys.\ Lett.}\  {\bf 115B}, 295 (1982).
  
\bibitem{Staro82} 
  	A. A. Starobinsky,
  	\emph{Phys.\ Lett.}\  {\bf 117B}, 175 (1982).

\bibitem{GuPi} 
	A. H. Guth, S. Pi, 
	\emph{Phys. Rev. Lett.} \textbf{49}, 1110 (1982). 

\bibitem{scalar} 
	A. R. Liddle, D. H. Lyth, 
	\emph{Phys. Lett. B} \textbf{291}, 391 (1992). 

\bibitem{Appleby} 
  	S. A. Appleby, R. A. Battye, A. A. Starobinsky,
  	\emph{JCAP} {\bf 1006}, 005 (2010).
  
\bibitem{HZS} 
	E. R. Harrison, 
	\emph{Phys. Rev. D} \textbf{1}, 2726 (1970). 

\bibitem{MSZ} 
	R. Myrzakulov, L. Sebastiani, S. Zerbini,
	\emph{Eur. Phys. J. C} \textbf{75}, 215 (2015). 

\bibitem{SM} 
	L. Sebastiani, R. Myrzakulov, 
	\emph{Int. J. Geom. Methods Mod. Phys.} \textbf{12}, 1530003 (2015). 

\bibitem{LINCHAOS} 
	A. D. Linde, 
	\emph{Phys. Lett. B} \textbf{129}, 177 (1983). 

\bibitem{RVZ23} 
	M. Rinaldi, G. Cognola, L. Vanzo, S. Zerbini, 
	\emph{Phys. Rev. D} \textbf{91}, 123527 (2015). 

\bibitem{RIN} 
	M. Rinaldi, 
	lecture presented at \emph{14th Marcel Grossmann Meeting on Recent Developments in Theoretical and Experimental General Relativity, Astrophysics, and Relativistic Field Theories}, 12--18 Jul, Rome, Italy (2015). 

\bibitem{Venturi} 
  	F. Cooper, G. Venturi,
  	\emph{Phys.\ Rev.\ D} {\bf 24}, 3338 (1981).

\bibitem{Turchetti} 
  	G. Turchetti, G. Venturi,
  	\emph{Nuovo Cim.\ A} {\bf 66}, 221 (1981).



\bibitem{Odintsov20}
S.~D.~Odintsov and V.~K.~Oikonomou,
Phys. Lett. B \textbf{807} (2020), 135576;

S.~D.~Odintsov and V.~K.~Oikonomou,
Phys. Lett. B \textbf{833} (2022), 137353.





\end{thebibliography}
\end{document}